\documentstyle[twoside,fleqn,espcrc2]{article}

\newcommand{\AmS}{{\protect\the\textfont2
  A\kern-.1667em\lower.5ex\hbox{M}\kern-.125emS}}

\title{\begin{flushright}
\footnotesize{ULB-TH-99-31/}\\
\footnotesize{December 1999}
\end{flushright}Symmetry-deforming interactions of chiral $p$-forms}

\author{X. Bekaert\address{Physique 
Th\'eorique et Math\'ematique, Universit\'e
Libre de Bruxelles, \\
Campus Plaine C.P. 231, B-1050 Bruxelles, Belgium}, M. Henneaux$^{\hbox{
\footnotesize{a}}}$%
        \thanks{Also at Centro de Estudios Cient\'{\i}ficos
        de Santiago, Casilla 16443, Santiago 9, Chile}
and
A. Sevrin\address{Theoretische Natuurkunde, Vrije Universiteit Brussel \\
Pleinlaan 2, B-1050 Brussel, Belgium}}
       
\begin{document}

\begin{abstract}
No-go theorems on gauge-symmetry-deforming interactions of chiral
$p$-forms are reviewed.  We consider the explicit case of $p=4$, 
$D=10$ and show that the only symmetry-deforming consistent vertex for a
system of one chiral $4$-form and two $2$-forms is the one that occurs 
in the type $II_B$ supergravity Lagrangian.
Article based on a talk given by M. H. at the conference ``Constrained
Dynamics and Quantum
Gravity 99" held in Villasimius (Sardinia), September 13-17 1999,
to appear in the proceedings of the meeting (Nucl. Phys. B Proc. Suppl.).
\end{abstract}

\maketitle

\section{INTRODUCTION}

Chiral $p$-forms, which can be defined in ($2p+2$)-dimensional
Minkowski spacetime for any even $p$,
have the puzzling feature of not admitting a simple, manifestly
covariant action principle even though their equations of
motion are Lorentz-invariant, and even though one can define
Poincar\'e generators that appropriately close according to
the Poincar\'e algebra.  The case $p=0$ (chiral bosons) arises in 
one formulation of the heterotic
string, while the cases $p=2$ and $p=4$ are relevant to 
the $M$-theory $5$-brane and type $II_B$-supergravity,
respectively.

Although there is no simple, manifestly covariant action principle, there
exists a non-manifestly covariant Lagrangian, which is quadratic in
the fields in the free case and  which yields the correct
dynamics \cite{HT}.  There exists also a non-polynomial,
manifestly covariant action \cite{Pasti}.  The two formulations
are equivalent since one goes from the second one
to the first one by appropriately gauge-fixing the auxiliary
pure gauge field introduced in \cite{Pasti}.

In a recent paper \cite{BHS}, we have shown that the local interactions
of chiral $p$-forms are severely constrained by the
consistency requirement that they should not modify the number of
physical degrees of freedom.  More precisely, if one imposes that
the deformed action (free action + interaction terms) be invariant
under a deformed set of local gauge symmetries that continously reduce
to the gauge symmetries of the free theory as the coupling goes to zero,
one finds in fact that no consistent interaction can deform
the gauge symmetry at all.  The gauge transformations must retain 
their original form under consistent deformations of the
action.  There is in particular no analog of the
non-abelian Yang-Mills construction for chiral $p$-forms.  The
only available interactions must be invariant under the gauge
symmetries of the abelian theory and leave the gauge
structure untouched. This
result generalizes to chiral $p$-forms the no-go theorems established
for non-chiral $p$-forms in \cite{RN,CT,Deetal,MH,HK} 

Motivated by the M-theory $5$-brane, we considered in \cite{BHS} the
explicit case of $p=2$.  However, the argument is clearly quite general
and applies to any $p$.   Although we used as starting point the 
non-manifestly covariant Lagrangian of \cite{HT}, the same obstructions
would be present had we worked in the ``PST formalism" of
\cite{Pasti} because, as we have recalled,
one recovers \cite{HT} from \cite{Pasti} by
appropriately fixing the gauge.  Furthermore, since the obstructions arise
already at the level of the gauge symmetries
(and not Lorentz invariance), they are easily detected in the
non-manifestly covariant approach.
 
Every no-go theorem has of course the weaknesses of its hypotheses.
In our case, these are:
\begin{enumerate}
\item The deformed action is local
\item The deformation is continuous
\item The system of local fields being dealt with is a system
of $N$ chiral $p$-forms described in the free limit by
the non-covariant action of \cite{HT} (or the covariant action
of \cite{Pasti}, see above). 
Scalar fields or spinor fields 
are also allowed. 
\end{enumerate}
We stress, in particular, that no restriction on the number $N$
of chiral $p$-forms was ever imposed. So, the $p$-forms can be labelled
by any number of indices. Furthermore, no assumption on the order
of the coupling vertices was made;  these are not necessarily
cubic, but can be quartic, quintic etc.

Accordingly, an interacting theory involving chiral $p$-forms with a deformed
gauge symmetry must be either non-local (which is the most likely possibility
in the M-theory $5$-brane case, perhaps in the context of
gerbes \cite{Kalk,Hitchin}), or non-perturbative and non-continuously
connected with the free case, or have a richer field content.

Even this last possibility appears to be rather restrictive, however, 
and subject to strong no-go theorems that forbid a non-abelian
deformation of the chiral $p$-form gauge symmetries 
analogous to the Yang-Mills deformation.  One finds
that the interactions may 
deform the gauge transformations but, when they do so,
they cannot deform their algebra (to
first-order in the deformation parameter). Furthermore,
those that do deform
the gauge transformations are few in number (if there is any at all).
All other consistent interactions are again off-shell gauge-invariant
under the abelian gauge symmetry and so do not deform it.
We have not checked the rigidity of the gauge algebra
in full generality in the chiral case, but the 
study of various explicit examples 
and the similarity with the non-chiral
case where it has been established \cite{HK} make us confident that
this rigidity holds in the same manner in
the chiral case. This leaves
only the first two possibilities for non-abelian deformations.

We illustrate in these proceedings the no-go theorems on the
symmetry-deforming interactions of exterior form gauge fields
(involving chiral ones)
by considering the system consisting of
a chiral $4$-form in ten spacetime dimensions together with $n$ 
$2$-forms.  This is relevant to type $II_B$-supergravity. 
We show that the only allowed symmetry-deforming
interactions are in fact precisely those
that appear in the supergravity Lagrangian.  
There are no others.  These interactions
deform the gauge transformations of the exterior forms but not their
algebra, which remain abelian.

This case is a good example of the general cohomological
techniques involved in the calculation of the
deformations.

\section{FIRST-ORDER DEFORMATIONS}

\subsection{Free action}

We start with the free action
\begin{equation}
S[A_{ijkl},B^a_{\lambda \mu}]= S_A + S_B
\label{freeaction}
\end{equation}
with
\begin{eqnarray}
S_A = & &\int d^{10}x (\frac{1}{10}
\varepsilon^{i_1 \dots i_5 j_1 \dots j_4}
F_{i_1 \dots i_5} \dot{A}_{j_1 \dots j_4}
\nonumber \\
&  &- \frac{4!}{10}
F^{i_1 \dots i_5}
F_{i_1 \dots i_5})
\label{freeactionA}
\end{eqnarray}
and
\begin{equation}
S_B = - \sum_a \frac{1}{2 \cdot 3!}
\int d^{10}x H^a_{\lambda \mu \nu} H^{a \lambda \mu \nu}.
\label{freeactionB}
\end{equation}
We have set
\begin{equation}
F_{ijklm} = 5 \, \partial_{[i} A_{jklm]}, \; \;  H^a_{\lambda \mu \nu} =
3 \, \partial_{[\lambda} B^a_{\mu \nu]}. 
\end{equation}
The PST manifestly covariant version of the theory
may be found in \cite{Dall1,Dall2}.
We leave the number of $2$-forms unspecified at this stage.  We shall see that
consistency to second-order forces the chiral $4$-form to interact
with only two $2$-forms.

The action is invariant under the following gauge transformations
\begin{equation}
\delta_{\Lambda,\epsilon} A_{j_1 \dots j_4} = 4 \partial_{[j_1} 
\Lambda_{j_2 j_3 j_4]}
\label{gaugeA}
\end{equation}
and
\begin{equation}
\delta_{\Lambda,\epsilon} B^a_{\lambda \mu} = \partial_{[\lambda}
\epsilon^a_{\mu]}
\label{gaugeB}
\end{equation}
The invariance of (\ref{freeactionB}) and of the
energy-term of (\ref{freeactionA})
is obvious; the invariance
of the kinetic term of (\ref{freeactionA})
follows from an integration by parts and the Bianchi identity
for the curvature $F_{i_1 \dots i_5}$.  The kinetic term
behaves thus like a Chern-Simons term.

The equations of motion that follow from (\ref{freeaction}) are
the standard ones for the $2$-forms and
\begin{equation}
4! \, \partial_m F^{m j_1 \dots j_4} 
- \varepsilon^{i_1 \dots i_5 j_1 \dots j_4}
\partial_{[i_1} \dot{A}_{i_2 \dots i_5]} = 0
\end{equation}
for the chiral $4$-form.  From this equation, one derives
\begin{equation}
\dot{A}_{i_1 \dots i_4} - \frac{1}{5!} 
\varepsilon_{i_1 \dots i_4 j_1 \dots j_5} F^{j_1 \dots j_5}
= 4 \, \partial_{[i_1} u_{i_2 i_3 i_4]}
\end{equation}
(assuming the $4$th Betti number of the spatial
sections to vanish).  This is the self-duality
condition $F = \, ^{*}\! F$ if one identifies the arbitrary
functions $u_{i_2 i_3 i_4}$ with the gauge components
$A_{0 i_2 i_3 i_4}$.

A chiral $4$-form and two $2$-forms are present in the spectrum of
type $II_B$ supergravity in $10$ dimensions
\cite{Schwarz1,Schwarz2,Howe}. The 4-form signals the presence of the
D3 brane and the two 2-forms correspond to the fundamental string and
the D1-brane, respectively.

\subsection{BRST differential}

We shall here just outline the general method and ideas
without going into the details.  These will be given in
\cite{BHS2}.

To determine the possible consistent interactions, we 
follow the method of \cite{BarH}.  The question boils down
to computing the BRST cohomological group $H^0(s \vert d)$ at
ghost number zero, i.e., one must find the general solution
of the cocycle condition
\begin{equation}
sa + db = 0, \; \; gh(a) =0
\label{WZ}
\end{equation}
modulo trivial ones (of the form $sm +dn$).
Here $s$ is the BRST differential and $d$ the spacetime exterior 
derivative.

The solutions of (\ref{WZ}) are expanded according to the
antighost (or antifield) number,
\begin{equation}
a = a_0 + a_1 + \cdots + a_k.
\label{expan}
\end{equation}
The fact that the expansion stops at a finite order $k$ follows from the fact
$a_0$ contains only a finite number of derivatives, which
already excludes a weak form of non-locality.

Each term in the expansion (\ref{expan}) has total ghost number zero.  The
first term does not involve the ghosts or the antifields.
The next terms depend both on the antifields and the ghosts, but
in such a way that the antighost number exactly balances the
pure ghost number (the higher $i$, the ``more" $a_i$
involves antifields and hence also ghosts).
The antifield-independent term $a_0$ defines 
the first-order consistent vertex.
The antifield-dependent terms define the deformation of the
gauge-symmetry and its algebra.  In particular,
the term $a_1$,
which is linear in the antifields conjugate to the classical
fields, defines the deformation of the
gauge transformations.  Similarly, the term in $a_2$ which is
linear in the antifields
conjugate to the first ghosts and quadratic in those ghosts, corresponds 
to the deformation of the gauge algebra.  If this term is absent,
the gauge algebra remains abelian (to first order) even after
the interaction is switched on.  This occurs if the most general
solution of (\ref{WZ}) is at most linear in the ghosts,
up to trivial terms that can absorbed through redefinitions.
This what happens for the model at hand.

The BRST differential is explicitly given by
$s = \delta + \gamma$
where one has, in the $4$-form sector with ghosts $C_{ijk}$,
ghosts of ghosts $D_{ij}$, $E_i$ and $\eta$, and antifields
$A^{*ijkl}$, $C^{*ijk}$, $D^{*ij}$, $E^{*i}$, $\eta^{*}$,
\begin{eqnarray}
\gamma A_{ijkl} &=& 4 \partial_{[i} C_{jkl]}, \; \; 
\delta A_{ijkl} = 0 \\
\gamma C_{ijk} &=& 3 \partial_{[i} D_{jk]}, \; \;
\delta C_{ijk} = 0 \\
\gamma D_{ij} &=& 2 \partial_{[i} E_{j]}, \; \;
\delta D_{ij} = 0 \\
\gamma E_i &=& \partial_i \eta, \; \;
\delta E_i = 0 \\
\gamma \eta &=& 0 , \; \;
\delta \eta = 0
\end{eqnarray}
and
\begin{eqnarray}
\delta A^{*ijkl} &= & 4! \, \partial_m F^{mijkl} \nonumber \\ & &- \,
\varepsilon^{ijklm_1 \dots m_5} \partial_{[m_1} \dot{A}_{m_2 \dots m_5]}, \\
\gamma  A^{*ijkl} &=& 0 \\
\delta C^{*ijk} &=& \partial_m A^{*mijk} , \; \; \gamma C^{*ijk} = 0, \\
\delta D^{*ij} &=& \partial_m C^{* mij} , \; \; \gamma D^{*ij} = 0, \\
\delta E^{*i} &=& \partial_m D^{*mi} , \; \; \gamma E^{*i} = 0, \\
\delta \eta^{*} &=& \partial_m E^{*m} , \; \; \gamma \eta^{*} = 0.
\end{eqnarray}
The ghosts $C_{ijk}$ and ghosts of ghosts $D_{ij}$, $E_i$ and $\eta$
have respectively pure ghost number $1$, $2$, $3$ and $4$.  The antifields
$A^{*ijkl}$, $C^{*ijk}$, $D^{*ij}$, $E^{*i}$ and $\eta^{*}$
have respectively antighost number $1$, $2$, $3$, $4$ and $5$.

In the $2$-form sector, the differentials $\delta$ and $\gamma$ are
given by
\begin{eqnarray}
\gamma B^a_{\lambda \mu} &=& \partial_\lambda \xi^a_\mu -
\partial_\mu \xi^a_\lambda, \; \; 
\delta B^a_{\lambda \mu} = 0, \\
\gamma \xi^a_\lambda &=& \partial_\lambda \sigma^a, \; \;
\delta \xi^a_\lambda = 0, \\
\gamma \sigma^a &=& 0, \; \; \delta \sigma^a = 0, \\
\gamma B_a^{* \lambda \mu} &=& 0 , \; \;
\delta B_a^{* \lambda \mu} = \partial_\rho H_a^{\rho \lambda \mu}, \\
\gamma \xi_a^{* \lambda} &=& 0, \; \;
\delta \xi_a^{* \lambda} = \partial_\rho B_a^{* \rho \lambda}, \\
\gamma \sigma_a^{*} &=& 0, \; \;
\delta \sigma_a^{*} = \partial_\rho \xi_a^{* \rho}.
\end{eqnarray} 
The ghosts $\xi^a_\lambda$ have pure ghost number $1$ and the ghosts of
ghosts $\sigma^a$ have pure ghost number $2$.  The antifields
$B_a^{* \lambda \mu}$, $\xi_a^{* \lambda}$ and $\sigma_a^{*}$
have respectively antighost number $1$, $2$ and $3$.

The (total) ghost number $gh$ is the difference between the pure ghost number
and the antighost number. The component $\delta$ and $\gamma$ of $s$ have
(pureghost,antighost) number equal to (0,-1) and (1,0) respectively. The equation
$s^2=0$ is equivalent to $\delta^2=\delta\gamma+\gamma\delta=\gamma^2=0$.  

\subsection{Deformations}
We assume $k>0$ in (\ref{expan}) since when $a$ reduces to
$a_0$, it does not modify the gauge transformations
and there is nothing to be demonstrated.  The idea of the proof
is to show that the building blocks out of which $a_k$ can be
made for $k>0$, have ghost numbers that cannot generically add up to zero.
Thus, there is no or very few ways to write down an acceptable $a_k$
for $k>0$.

The building blocks of $a_k$ are, on the one hand, 
the polynomials in the last ghosts of ghosts $\sigma^a$ and
$\eta$ and their temporal derivatives $\partial_0 \eta$,
$\partial_{00} \eta$, etc, and, on the other hand, 
the elements of the ``invariant characteristic
cohomology" $H_k(\delta \vert d)$ or $H(\delta \vert \tilde{d})$,
where $\tilde{d}$ is the spatial exterior derivative $dx^i \partial_i$.
The former have even ghost number, while the latter generically have odd
antighost number.  So, their products generically fail to have
total ghost number zero, with only one exception for the
system under study.

To see this, one follows
the argument given in \cite{BHS} and one finds that the last
term $a_k$ in the general solution of the cocycle condition 
(\ref{WZ}) can be assumed to be annihilated by $\gamma$,
$\gamma a_k = 0$.  Thus, it reads
\begin{equation}
a_k = \sum_I P^I \omega^I
\label{sum}
\end{equation}
(up to trivial terms), where $P^I$ is a polynomial in the curvatures
$F_{ijklm}$ and $H^a_{\lambda \mu \nu}$, in the antifields, and in their
spacetime derivatives, while the $\omega^I$ form a basis in the algebra
of polynomials in the variables $\sigma^a$,
$\eta$, $\partial_0 \eta$, $\partial_{00} \eta$, etc, which are
the generators of the cohomology $H(\gamma)$ in positive ghost number [the fact that the time derivative of eta is
a generator  while the time derivative of sigma isn't is a consequence from
the fact that the 4-form is treated non-covariantly while the 2 form is
treated covariantly].

The coefficient $P^I$ of $\omega^I$ in the sum (\ref{sum}), which
is a $10$-form,
is subject to different conditions depending on to whether the
element $\omega^I$ that it multiplies depends on $\eta$ and its time derivatives
or not.
\begin{enumerate}
\item If $\omega^I$ involves at least one time derivatives of $\eta$
(including $\partial_0^0 \eta \equiv \eta$), then the
corresponding coefficient $P^I$ must be of the form
$P^I = Q^I dx^0$ where $Q^I$ is a spatial $9$-form
solution of $\delta Q^I + \tilde{d} N^I = 0$. Here, $\tilde{d}$
is the spatial exterior derivative, $\tilde{d} a = dx^k \partial_k a$.
However, there is no non-trivial solution to that equation that would
match the ghost number of $\omega^I$, which is an even number $\geq 4$
since $\omega^I$ contains at least one $\eta$ or one of its time derivatives.
The only non-trivial solutions are indeed in antighost number $5$, $2$ and $1$.
Thus, in this case, 
one cannot construct an $a_k$ ($k>0$)
with total ghost number zero.
This is exactly the situation described in
\cite{BHS} - and is the only case to be considered for a pure
system of chiral $p$-forms. Note the similarity with the
Hamiltonian analysis of consistent couplings done in
\cite{Bizda}. 
\item If $\omega^I$ depends only on $\sigma^a$, the corresponding
coefficient $P^I$ must be a $10$-form that solves the equation 
$\delta P^I + dN^I = 0$.  It must be of even antighost number to
match the even ghost number of $\omega^I$.  There is one possibility,
namely, $C^{*} \wedge H^a$, where $C^{*}$ is the $7$-form
$\varepsilon_{0 i_1 \dots i_6 j_1 j_2 j_3} C^{* j_1 j_2 j_3}
dx^0\wedge dx^{i_1} \wedge \dots \wedge dx^{i_6}$.  It has
antighost number $2$ and can be combined with $\sigma^a$ to yield
a non-trivial $a_2$.  Specifically - and in density notations -,
\begin{equation}
a_2= C^{*i_1 i_2 i_3}H^a_{i_1 i_2 i_3} \sigma^b \mu_{ab}
\end{equation}
where $\mu_{ab}$ is a matrix of coupling constants having
dimension (length)$^4$.  The matrix
$\mu_{ab}$ must be antisymmetric since otherwise
$H^a_{i_1 i_2 i_3} \sigma^b \mu_{ab} = \gamma$(something)
$+ d$(something) and $a_2$ can be removed.
This possibility specifically requires the presence of the $2$-forms
and is not available for a pure system of chiral forms.
Besides $C^{*} \wedge H^a$,
there is no other non trivial class in $H_k(\delta \vert d)$
with even antighost number.
\end{enumerate}

Since the above $a_2$ is the only possibility, we carry on the
discussion
with it.
The equation $\delta a_2 + \gamma a_1 + db_1 = 0$
determines $a_1$ up to a trivial term,
\begin{equation}
a_1 = - A^{*m i_1 i_2 i_3} H^a_{i_1 i_2 i_3} \xi^b_m \mu_{ab}. 
\end{equation}
The corresponding interaction vertex $a_0$ is then found to be 
\begin{eqnarray}
a_0= &-& 12 F^{kmi_1i_2i_3}H^a_{i_1 i_2 i_3} B^{b}_{km} 
\mu_{ab} \nonumber \\
&-& \frac{3}{10} \varepsilon^{i_0 \dots i_3 l_1 \dots l_5}
F_{l_1 \dots l_5} H^a_{0 i_2 i_3} B^b_{i_1 i_0} \mu_{ab}
\nonumber \\
&+& \frac{1}{5} \varepsilon^{i_0 \dots i_3 l_1 \dots l_5}
F_{l_1 \dots l_5} H^a_{i_1 i_2 i_3} B^b_{0i_0} \mu_{ab}
\label{firstorder}
\end{eqnarray}
The term $a_0$ is determined up to vertices
that are gauge-invariant under the abelian gauge-symmetry
(possibly modulo a total derivative).  Since these do
not modify the gauge transformations, we do not include them
in the sequel
and stick to the choice (\ref{firstorder}) for $a_0$.
This choice has the smallest number of derivatives.

\section{COMPLETE ACTION}

\subsection{New gauge symmetries}
Once the first-order vertex is determined, one can analyse the conditions 
imposed by consistency to second-order.  These conditions have also a
cohomological interpretation \cite{BarH}.  We shall complete
the action in the explicit case where the first-order vertex is
solely given by the above deformation $a_0$, 
without extra gauge-invariant terms.

To discuss the consistency conditions at second-order,  we first note
that the gauge transformation of the field $A_{ijkl}$ is
modified by the interaction since $a$ contains a piece
linear in $A^{*m i_1 i_2 i_3}$, from which one reads the new gauge
transformations
\begin{equation}
\delta_{\Lambda, \epsilon} A_{ijkl} =
4 \, \partial_{[i} \Lambda_{jkl]} + \epsilon^a_{[i}H^b_{jkl]} 
\mu_{ab}
\label{newgauge}
\end{equation}  
The term $a_2$, linear in the antifield
$C^{*ijk}$, signals the corresponding
modification of the reducibility identity.
The gauge transformations of $B^a_{\mu \nu}$ are unchanged
since there is no term proportional to $B^{*\mu \nu}_a$ in $a_1$.

\subsection{Action}
To derive the complete action,
it is convenient
to introduce the redefined field strength
\begin{equation}
{\cal F}_{i_1 \dots i_5} = F_{i_1 \dots i_5} - \frac{5}{2}
B^c_{[i_1 i_2} H^d_{i_3 i_4 i_5]} \mu_{cd}
\end{equation}
which has the property of being invariant under the
transformation (\ref{newgauge}).
Similarly, one defines
\begin{eqnarray}
{\cal D}_0 A_{j_1 \dots j_4} = & & \dot{A}_{j_1 \dots j_4}
- \frac{3}{2} B^a_{[j_1j_2} H^b_{0j_3j_4]} \mu_{ab}
\nonumber \\
& & - B^a_{0[j_1} H^b_{j_2 j_3 j_4]} \mu_{ab}
\end{eqnarray}
which transforms into spatial total derivative, $\delta_{\Lambda, \epsilon}
{\cal D}_0 A_{j_1 \dots j_4} \sim \partial_{[j_1} f_{j_2 j_3 j_4]}$.

By following the recursive construction of \cite{BarH},
one finds that $O(\mu^2)$-terms must be added to the action.
There is no modification to the gauge transformations at that order,
which remain given by (\ref{newgauge}) and $\delta_{\Lambda, \epsilon}
B^a = d \epsilon^a$.
Rather than following in detail the recursive construction of \cite{BarH},
we shall give the final answer.  The complete action takes the form 
\begin{equation}
\tilde{S}[A_{ijkl}, B^a_{\lambda \mu}] = 
\tilde{S}_A + S_B, 
\label{complete}
\end{equation}
with $\tilde{S}_A$ given by
\begin{eqnarray}
\tilde{S}_A &=& \int d^{10}x ( \frac{1}{10} \varepsilon^{i_1 \dots i_5
j_1 \dots j_4} {\cal F}_{i_1 \dots i_5} {\cal D}_0 A_{j_1 \dots j_4}
\nonumber \\
& & - \frac{1}{2} \varepsilon^{i_1 \dots i_5 j_1 \dots j_4} 
H^a_{0i_1 i_2} H^b_{i_3 i_4 i_5} A_{j_1 \dots j_4} \mu_{ab} 
\nonumber \\
& & - \frac{4!}{10} {\cal F}^{i_1 \dots i_5} {\cal F}_{i_1 \dots i_5})
\end{eqnarray}
and $S_B$ unchanged.

One easily verifies that the complete action is invariant under the new
gauge symmetries provided the $10$-form 
\begin{equation}
\epsilon^a \wedge H^b \mu_{ab} \wedge H^c \wedge H^d \mu_{cd}
\label{condonmu}
\end{equation}
vanishes.
This condition will be fulfilled only if the antisymmetric
matrix $\mu_{ab}$ is of rank two.  This is most easily seen
by bringing $\mu_{ab}$ to canonical form,
\begin{equation}
\mu_{ab} =
\left( \begin{array}{lllllll}
0&\alpha&0&0&\dots&0&0 \\
-\alpha&0&0&0&\dots&0&0  \\
0&0&0&\beta&\dots&0&0 \\
0&0&-\beta&0&\dots&0&0\\
\cdot&\cdot&\cdot&\cdot&\dots&\cdot&\cdot \\
0&0&0&0& \dots& 0& \lambda \\
0&0&0&0& \dots&-\lambda&0
\end{array}
\right)
\label{matrix}
\end{equation}
where we have assumed an even number of $2$-forms.  [In the case
of an odd number of $2$ forms, there would be an extra row and an extra
column of zeros, making it clear that the ``last" $2$-form decouples.]
Only one eigenvalue can be non-zero in
(\ref{matrix}): if two are non-zero, say 
$\alpha$ and $\beta$, then the sum (\ref{condonmu})
contains the term $\alpha \beta \epsilon^1\wedge
H^2 \wedge H^3 \wedge H^4$, which is non-zero.
Thus we see that consistency to second order imposes
that only two $2$-forms couple in fact to the
chiral $4$-form. [The term (\ref{condonmu}), with
$\epsilon^a$ replaced by the ghost $\xi^a$, is
a non-trivial solution of $s a +db$ and so represents a true
obstruction, unless it is zero.]

The action (\ref{complete}) with two $2$-forms is the action
for the bosonic sector of type $II_B$-supergravity (with
scalar fields set equal to zero and the metric equal to $\eta_{\mu \nu}$). 

\subsection{Equations of motion}
The equations of motion that follow from varying the complete action
with respect to $A_{ijkl}$ are 
\begin{equation}
4! \, \partial_m {\cal F}^{mijkl} 
- \varepsilon^{m_1 \dots m_5 ijkl}
\partial_{[m_1} {\cal D}_0 A_{m_2 \dots m_5]} = 0
\end{equation}
Assuming as before that the 4th Betti number of the spatial
sections is zero, one gets
\begin{equation}
{\cal D}_0 A_{ijkl} - \frac{1}{5!} 
\varepsilon_{ijkl m_1 \dots m_5} {\cal F}^{m_1 \dots m_5}
= 4 \, \partial_{[i} u_{jkl]}.
\label{EqOfMo}
\end{equation}
Defining $A_{0ijk}$ to be $u_{ijk}$, one can rewrite
(\ref{EqOfMo}) as
\begin{equation}
{\cal F} = ^* \! \! {\cal F}
\end{equation}   
where
\begin{equation}
{\cal F} = dA - \frac{5}{2} \mu_{ab} B^a \wedge H^b.
\end{equation}
These are the modified chirality conditions in covariant form.

The equations of motion for the $2$-forms are, on the other hand
\begin{equation}
d ^*\! H^a + 3!\,5!\,\mu_{ab} H^b {\cal F}=0.
\end{equation}
Note that one could have derived the action (\ref{complete}) by
starting from the action describing a non-chiral $4$-form
interacting with two $2$-forms through Chapline-Manton and
Chern-Simons couplings with coefficients such that $d{\cal F} = d ^* \! {\cal F}$
and making the projection to the chiral sector through constraints.

\subsection{Algebra of new gauge transformations}
It is easy to verify that the algebra of the new gauge transformations
is abelian.  This is because the new terms involve only the curvatures
and no bare $A$ or $B^a$.  Thus, the gauge algebra is not deformed.

\section{CONCLUSIONS}

In this paper, we have studied the interactions for a mixed system
containing a chiral $4$-form and $n$ $2$-forms in $10$ dimensions.
We have shown the uniqueness of the supergravity vertices.
Any interaction that deforms the gauge transformations necessarily
involves them.
The gauge algebra is, however, untouched. We stress that this result has been obtained
without invoking supersymmetry. Our analysis illustrates
the rigidity of $p$-form systems in the local field-theoretical context,
already well appreciated in the non-chiral case.

\section*{ACKNOWLEDGEMENTS}
M.H. is grateful to the organizers of the meeting ``Constrained
Dynamics and Quantum Gravity 99"
for the very enjoyable conference in Sardinia.
X.B. and M.H. are suported in part by the ``Actions de
Recherche Concert{\'e}es" of the ``Direction de la Recherche
Scientifique - Communaut{\'e} Fran{\c c}aise de Belgique", by
IISN - Belgium (convention 4.4505.86) and by
Proyectos FONDECYT 1970151 and 7960001 (Chile).
A.S. is supported in part by the FWO and by the European
Commission TMR programme ERBFMRX-CT96-0045 in which he is
associated to K.\ U.\ Leuven.


\begin{thebibliography}{99}
\bibitem{HT}
M.~Henneaux and C.~Teitelboim,
``Dynamics Of Chiral (Selfdual) p-Forms,''
Phys.\ Lett.\ {\bf B206} (1988) 650.
\bibitem{Pasti} P.~Pasti, D.~Sorokin and M.~Tonin,
``On Lorentz invariant actions for chiral p-forms,''
Phys.\ Rev.\ {\bf D55} (1997) 6292
hep-th/9611100.
\bibitem{BHS} X. Bekaert, M. Henneaux and A. Sevrin,
``Deformations of chiral two-forms in six dimensions,''
Phys.\ Lett.\ {\bf B468} (1999) 228, hep-th/9909094.
\bibitem{RN} R.I.~Nepomechie,
``Approaches To A Nonabelian Antisymmetric Tensor Gauge Field Theory,''
Nucl.\ Phys.\ {\bf B212} (1983) 301.
\bibitem{CT} C.~Teitelboim,
``Gauge Invariance For Extended Objects,''
Phys.\ Lett.\ {\bf B167} (1986) 63.
\bibitem{Deetal} T.~Damour, S.~Deser and J.~McCarthy,
``Nonsymmetric gravity theories: Inconsistencies and a cure,''
Phys.\ Rev.\ {\bf D47} (1993) 1541,
gr-qc/9207003.
\bibitem{MH} M.~Henneaux,
``Uniqueness of the Freedman-Townsend Interaction Vertex For 
Two-Form Gauge Fields,''
Phys.\ Lett.\ {\bf B368} (1996) 83,
hep-th/9511145. 
\bibitem{HK} M.~Henneaux and B.~Knaepen,
``All consistent interactions for exterior form gauge fields,''
Phys.\ Rev.\ {\bf D56} (1997) 6076,
hep-th/9706119.
\bibitem{Kalk} J. Kalkkinen,
``Gerbes and massive type II configurations,'' 
JHEP {\bf 9907} (1999) 002,
hep-th/9905018; ``Non-Abelian gerbes from strings on a branched space-time,''
hep-th/9910048.
\bibitem{Hitchin} N. Hitchin, 
``Lectures on special Lagrangian submanifolds,''
math.DG/9907034.
\bibitem{Dall1} G.~Dall'Agata, K.~Lechner and D.~Sorokin,
``Covariant actions for the bosonic sector of D = 10 IIB supergravity,''
Class.\ Quant.\ Grav.\ {\bf 14} (1997) L195,
hep-th/9707044.
\bibitem{Dall2} G.~Dall'Agata, K.~Lechner and M.~Tonin,
``D = 10, N = IIB supergravity: Lorentz-invariant actions and duality,''
JHEP {\bf 07} (1998) 017,
hep-th/9806140.
\bibitem{Schwarz1} J.H.~Schwarz,
``Covariant Field Equations Of Chiral N=2 D = 10 Supergravity,''
Nucl.\ Phys.\ {\bf B226} (1983) 269.
\bibitem{Schwarz2} 
J.H.~Schwarz and P.C.~West,
``Symmetries And Transformations Of Chiral N=2 D = 10 Supergravity,''
Phys.\ Lett.\ {\bf B126} (1983) 301.
\bibitem{Howe}
P.S.~Howe and P.C.~West,
``The Complete N=2, D = 10 Supergravity,''
Nucl.\ Phys.\ {\bf B238} (1984) 181.
\bibitem{BHS2} X. Bekaert, M. Henneaux and A. Sevrin, in preparation.
\bibitem{BarH} G.~Barnich and M.~Henneaux,
``Consistent couplings between fields with a gauge freedom and deformations 
of the master equation,''
Phys.\ Lett.\ {\bf B311} (1993) 123,
hep-th/9304057.
\bibitem{Bizda} C. Bizdadea, L. Saliu and S. O. Saliu, 
``Chapline-Manton interaction vertices and Hamiltonian BRST
cohomology", Int. J. Mod. Phys. A, to appear.
\end{thebibliography}
\end{document}